# Intra-valence-band mixing in strain-compensated SiGe quantum wells

S. Tsujino[1,a], A. Borak[1], C. Falub[1], T. Fromherz[2], L. Diehl[1,b], H. Sigg[1], and D. Grützmacher[1]

[1]Laboratory for Micro- and Nanotechnology, Paul Scherrer Institut, CH-5232 Villigen-PSI, Switzerland

[2]Institut für Halbleiter- u. Festkörperphysik, Universität Linz, A-4040 Linz, Austria

6 April 2005

We explore the midinfrared absorption of strain-compensated $p$-$Si_{0.2}Ge_{0.8}$/Si quantum wells for various well thicknesses and temperatures. Owing to the large band offset due to the large bi-axial strain contrast between the wells and barriers, the intersubband transitions energies from the ground state to the excited heavy hole (hh), light hole (lh) and split-off hole (so) states are resolved to ~0.5 eV. When hh2 is within ~30 meV of lh1 or so1 a partial transfer of the hh1-hh2 oscillator strength to the hh1-lh1 or hh1-so1 transitions is observed, which is otherwise forbidden for light polarized perpendicular to the plane of the wells. This is a clear sign of mixing between the hh and lh or so-states. A large temperature induced broadening of hh2 peak is observed for narrow wells indicating a non-parabolic dispersion of the hh2 states due to the mixing with the lh/so continuum. We found that the 6-band **k·p** theory gives a quantitative account of the observations. A possible role of many-body effects in the temperature-induced negative peak shift is discussed.

73.21.Fg, 78.30.Hv, 78.67.De

---

[a] Electronic mail: soichiro.tsujino@psi.ch

[b] Present address: Div. of Engineering and Applied Sciences, Harvard university, Cambridge, MA 02138

A detailed understanding of the intra-valence band structure in semiconductor quantum wells (QWs) is important to optimize the performance of opto-electronic devices including the vertical incidence QW infrared detectors[1] and quantum cascade (QC) light emitters[2,3] based on intersubband transitions in a two-dimensional hole gas (2DHG). The energy levels in the valence band are largely influenced by the mixing between states originating from three bands with different angular momentum; the heavy hole (hh, |3/2, ±3/2>), light hole (lh, |3/2, ±1/2>), and spin-orbit split-off hole (so, |1/2, ±1/2>) band. Intersubband optical absorption employed here can detect the excited valence-subbands by optically exciting holes with an in-plane wave vector $k$ smaller than the Fermi wave vector $k_F$ and can provide useful data up to temperatures well above the room temperature.[4] Magneto-tunneling spectroscopy of double barrier structures can map out the highly non-parabolic hole dispersion,[5,6] however, in a 2DHG the resonant tunneling can be dominated by emitter states with $k$ equal to $k_F$,[7] or may fail to detect states with a different angular momentum from that of the emitter states when the mixing is sufficiently weak.[8]

In the past, the valence band structure in pseudo-morphic SiGe QWs has been studied in detail by midinfrared intersubband optical absorption.[9,10] Although the band parameters in strained SiGe with arbitrary Ge-content have been known theoretically for some time,[11] high Ge-content SiGe QWs have been explored experimentally by intersubband optical absorption only recently.[12] In strain-compensated $Si_{0.2}Ge_{0.8}$/Si QWs grown on $Si_{0.5}Ge_{0.5}$ virtual substrates, the valence band offset is equal to ~0.6 eV owing to the large strain-contrast of 3% between the wells and barriers. Therefore, band-edge hole states with different angular momenta can be energetically separated and the transition resonances are well resolved to ~0.5 eV. We show that by changing the well width to bring the energy separation between excited hh and lh or so states to within ~ 30 meV, the mixing between these states can be observed as a transfer of the oscillator strength in the polarization resolved absorption.

The strain-compensated QWs are grown by low-temperature molecular beam epitaxy on $Si_{0.5}Ge_{0.5}$ virtual substrates. The samples S1 to S6 consist of four periods of modulation doped $Si_{0.2}Ge_{0.8}$ wells sandwiched by Si barriers. Previous studies have shown that these samples have sharp and flat Si-$Si_{0.2}Ge_{0.8}$ interfaces characterized by an interface roughness height $\Delta$ of 0.4 nm and a correlation length $\Lambda$ of 2.3 nm.[13] The thicknesses of the QWs are, from S1 to S6, 2.5, 3.0, 3.5, 4.5, 5.5, and 7.0 nm respectively. The thicknesses of the Si barriers are 1.8 nm for the first four samples, 1.7 nm for S5, and 2.1 nm for S6. Carrier concentration of the 2DHG is nominally $1\times10^{12}$ $cm^{-2}$ per QW, so only the first hh subband (hh1) is occupied at low temperature. Detailed layer sequences and the growth procedure have been described previously.[13]

In the experiment, we measured the infrared absorption spectra by electrical modulation in a 45 degree multi-pass waveguide geometry using a Fourier transform infrared spectrometer in step-scan mode. When an alternating bias is applied between Ti Schottky electrodes and Al Ohmic contacts, the waveguide transmission is modulated by $\Delta T$ due to the modulation of the carrier concentration. The transmitted polarized infrared light $T$ is detected by a liquid nitrogen cooled mercury-cadmium-telluride detector or a liquid Helium cooled Si-bolometer, and the $\Delta T$ component separated by lock-in techniques. Finally the single-pass absorption spectrum is obtained for each sample from the ratio $\Delta T/T$ divided by $l/(2t)$ where $l$ and $t$ are length and thickness of the waveguide, respectively. The gate bias is alternated between 2.0 V and –0.5 V. The bias swing corresponds to a carrier modulation of ~$1.0 \times 10^{12}$ cm$^{-2}$, calculated using the capacitance of the samples. In our measurement, the optical electric field is in the plane of the QWs for transverse electric (TE) polarization. In transverse magnetic (TM) polarization, the optical electric field has both in-plane and vertical components. However, we can identify the polarizations of the transitions by comparing the absorption intensities of the TM- and TE-absorption since the parallel absorption component in the TM-geometry is approximately equal to one half of the TE-absorption in our geometry.

We show in Figure 1(a) the single-pass absorption spectra of the samples measured at 15 K. The strong absorption lines in the TM polarized spectra are due to intersubband transitions from hh1 to hh2. As the well thickness decreases, the hh2-peak shifts to higher frequency and the linewidth is increased. In TE polarization, the hh1-lh2 transition is observed on the high frequency side of the hh2 absorption. The TE polarized peaks around 100 meV are ascribed to hh1-lh1 transitions (see below). Additionally to these, a broad absorption band is observed above ~350 meV due to the transitions from hh1 to so-states mixed with lh continuum states.[12]

For S5 and S6, the hh2 and lh1 peaks are observed as doublets at around 100 meV both in the TM and TE, indicating that these states are strongly mixed in these samples. For S6 in TM, the doublet consists of two almost equally strong peaks, although, as will be argued below, the peak at 108 meV can be identified as the hh1 to lh1 transition which normally is detected only by in-plane polarized light.[9,10,14] Here, the TE absorption of this transition is even smaller by a factor of 10 than the corresponding absorption in TM. Furthermore, the integrated absorption of the hh2/lh1 doublets in S5 and S6 is equal to those of the hh1-hh2 peak of the other samples within the experimental accuracy of ~10%. Therefore we conclude that the oscillator strength of the vertically polarized hh1-hh2 transition is partly transferred to the hh1-lh1 transition by the mixing between the hh2 and lh1 states. The mixing here is much stronger than in previous works[9,10,14] because of the small separation between the hh2 and the lh1 states.

For comparison, we calculated the band-edge intersubband transition energies using the 6-band **k•p** method but assuming $k$ equal to zero and solving the Schrödinger equation self-consistently [Fig.1(b) and Fig.2(b)]. We found that the calculated peak position energies agree reasonably well with the experiment as shown in Fig.1 (b), confirming the peak assignments. In the calculation, we included the exchange-correlation (XC) effect within the local density functional approximation (LDA). The depolarization shift is neglected since it is calculated to be below 3 meV in our samples. We used strain-dependent band offsets and Luttinger parameters evaluated by the method described in Ref. 11, 15, and 16. At $k = 0$, opposite spin states are degenerate and the hh states are decoupled from the lh/so states. We found that the inclusion of the lh-so coupling should be taken into account even at $k = 0$ to describe the lh/so peaks quantitatively, since the strain strongly couples the lh/so-states. In strained $Si_{0.2}Ge_{0.8}$ on $Si_{0.5}Ge_{0.5}$ substrates, lh-states has 25 % so-character. The mixing is further enhanced for the QW states that have a larger vertical momentum, although the states relevant to the understanding of the absorption spectrum still retain their original character in the energy range considered here.

From the comparison between the experiment and the calculation, we identify the shoulders observed in the low frequency tails of TE-polarized lh2-peaks for S1 and S3 as the overlapping broad peaks from hh1-so1 transitions and in-plane polarized hh1-hh2 transitions. A part of the low energy tail is likely due to the tunneling-induced coupling of lh2 states to the unstrained bulk $Si_{0.5}Ge_{0.5}$ states outside the QWs in these samples [Fig. 1(a) and Fig.2(b)]. Also, we ascribe the weak TE-polarized hh1-hh2 absorption in S1 and S3 to the mixing of the hh2-state with the lh2-state and/or so1 state as well as with the bulk $Si_{0.5}Ge_{0.5}$ states at finite $k$.

The $k$-dependent coupling between the hh and lh/so states are described by off-diagonal terms in the Luttinger-Kohn (LK) Hamiltonian.[17] Therefore, the mixing effect is enhanced at increased temperature since the high-$k$ states are occupied. We show in Fig.2(a) the temperature dependent absorption spectra of S3. Here, the 52 meV-separation between hh2 and so1 is comparable to the separation between the hh2-lh1 doublets in S5 and S6 but the absorption spectrum at the lowest temperature did not show a clear TM-polarized so1-absorption. However, when the temperature is increased above 160 K [Fig.2] and the hole population is enhanced for $k$-states above $k_F$, the TM-polarized absorption spectra develops a broad peak in the high frequency side of the hh2 peak. The peak intensity increases steadily to the temperature and the center of the peak coincides with the peak position energy of the TE-polarized so1 absorption. Since the TE-polarized absorption spectrum at high temperature does not show a comparable increase of the so1 absorption, we conclude that hh2 states with $k$ larger than $k_F$ are coupled with so1 states and the vertical transition oscillator strength is partially transferred from hh1-hh2 to hh1-so1 for those high $k$-states. A similar mixing effect is not observed in S4 in the measured temperature range between 15 K and 300 K [Fig. 2(a)] where the energy separation of hh2 state from lh1/so1 state is larger than in all the other samples. We note that the hh1-

hh2 absorption spectrum of S3 shows a considerable broadening from 19 meV at 15 K to 31 meV at 300 K in half width at the half maximum in the low frequency side but the linewidth of S4 increased by only 2.8 meV. This factor of 4.5 larger broadening of S3 compared to S4 is likely due to the mixing of the hh2 state with lh/so continuum at high-$k$. The small linewidth increase of S4 indicates that non-parabolicity is relatively small because hh2 states is lower than the $Si_{0.5}Ge_{0.5}$ band edge and therefore does not mix with lh/so continuum states for this sample [Fig.2(b)].

We evaluated the relevant off-diagonal terms of the LK Hamiltonian by using the wave functions calculated at $k = 0$. The hh-lh mixing between different parity states occurs via a term $L$ proportional to $|k|\langle k_z \rangle$, where $\langle k_z \rangle$ is the matrix element of the wave vector $k_z$ in the direction perpendicular to the wells. When the two unperturbed states have equal energies, these couple and anti-cross with the splitting energy given by $2|L|$. We found that $2|L|$ averaged over $k_F$ is equal to ~30 meV for S5 and S6. From the observed hh2-lh1 separation and the absorption ratio, we evaluated the anti-crossing energy and the hh2-lh1 separation $\delta E^{(0)}$ of unperturbed states. For S6, the two peaks have approximately equal strength suggesting that hh2 and lh1 are in resonance [$\delta E^{(0)} \sim 0$]. Therefore, $2|L|$ is approximately equal to the observed hh2-lh1 separation of 28 meV and agrees with the theoretical value. The resonance condition $\delta E^{(0)} = 0$ is predicted at a smaller well width than observed which is ascribed to the deviation of the exact well thickness and/or the Ge content from the designed values due to the growth uncertainty and to the uncertainty of the band parameters. A similar consideration applied to S5 yields values of $2|L|$ of 30 meV, and $\delta E^{(0)}$ of 24 meV, in good agreement with the calculated values. We note that the same mechanism leads to the observation of the parity forbidden interband excitonic transition between hh2 and and the first electron subband in GaAs QWs.[18,19] The effect is by a factor of 3-5 larger in our SiGe QWs because of the smaller well thicknesses and the larger $k$ range involved in the process.

In wells having thicknesses of ~3nm, the strongest coupling occurs between hh2 and so1 via the off-diagonal term $|iL/\sqrt{2}|$. A similar calculation as in the hh2-lh1 case gives the hh2-so1 splitting energy equal to 45 meV for a wave vector of 1.4 $k_F$, given by the thermal spread at 160 K.[20] The value is comparable to the hh2-so1 separation of 52 meV in S3 and therefore is consistent with the observed TM polarized so1 absorption at high temperatures.

Figure 3(b) shows the absorption coefficient of S3 at 15 K and 200 K calculated by the $k$-dependent 6-band **k·p** model. An excellent overall agreement with experiment is achieved. The calculation reproduces the increased TM polarized so1 peak at ~280 meV at high temperatures, the temperature-induced broadening and the low frequency peak shift of the hh2 peak but the predicted peak shift is a factor of 2 smaller than in the experiment. We note that, since the temperature dependence and the dispersion of the XC energy $E_{xc}$ are difficult to estimate by LDA employed here, the exact contribution of the many-body effect is still unclear but can be responsible for the low frequency shift of the hh2 peaks at increased temperature. We observed a 8.6 meV shift for S3 and a 6.0 meV shift for S4 when the temperature is increased from 15 K to 300 K [Fig. 2(a)]. Although the 2.6 meV larger shift in S3 reflects the stronger non-parabolicity and mixing with lh/so states of the hh2 state, the similar peak shift values of the two samples suggests the importance of the temperature dependence of $E_{xc}$ as in the case of the intersubband transitions in $n$-GaAs QWs[21] and $n$-InAs QWs.[22] Similarly to these QWs, we found that peak shift due to the temperature dependence of the band parameters is one order of magnitude too small and cannot account for the experiment in *p*-SiGe QWs.

Finally we discuss the well-thickness dependence of the broadening of the hh1-hh2 transitions that are relevant to the optical gain in SiGe QC light emitters.[3] The rapid reduction of the linewidth of the hh2 peak with the decrease of well-thickness suggests the importance of the interface-roughness-induced broadening.[23] However, *k*-dependent effects have a large influence on the broadening in our QWs with finite carrier population. We note that the hh-lh/so mixing can reduce the roughness-induced broadening of the hh1-hh2 absorption since the lh/so band offsets are smaller by a factor of 2 than the hh band offset. Also the non-parabolicity of the hh2 subband caused by the hh-lh/so mixing can be equally important. For narrow QWs below ~4 nm, the mixing effect is further enhanced by the coupling to the bulk $Si_{0.5}Ge_{0.5}$ states [Fig. 2]. A theory of the linewidth broadening of hh transitions which takes into account both these effects has not been reported yet.

The authours acknowledge Y. Campidelli, O. Kermarrec, and D. Bensahel of STMicroelectronics for providing the SiGe virtual substrates, A. Weber for her help with the device processing, S. Stutz for his help with the sample preparation, and U. Gennser for discussions. The work has been partially supported by the Swiss National Foundation and the European community within the SHINE project.

---

**Figure 1**

(a) TM- and TE-polarized intersubband absorption spectra of the strain-compensated $Si_{0.2}Ge_{0.8}$ QWs at 15 K for well thicknesses between 2.5 nm and 7.0 nm. The hole transitions from hh1 state to hh2, lh1, lh2, so1, and so2 states are indicated. Each spectrum is normalized by the dimension of the sample and represents the single-pass absorption. (b) Summary of the peak position energies. Lines show transition energies calculated by a self-consistent 6-band **k•p** model at the band edge ($k = 0$).

**Figure 2**

(a) Tempeature dependence of the TM-polarized hh1-hh2 absorption spectrum of S3 (right) and S4 (left). Each spectrum is normalized by the peak intensity and shifted horizontally for clarity. Measurement temperature is changed from 300 K to 15 K with 20 K steps from the top to the bottom except for the lowest temperature. (b) Schematic self-consistently calculated valence band structures and wave functions of $Si_{0.2}Ge_{0.8}$/Si QWs. The top and the bottom panel show the hh-band profiles and QW state wave functions for S3 and S4, respectively. The hh2 states of S3 and wells below ~4 nm are coupled to the continuum in $Si_{0.5}Ge_{0.5}$ layers outside of the QWs (not shown) by tunneling through the Si barriers.

**Figure 3**

(a) Polarized absorption spectra of S3 measured at 15 K and 200 K. The top panel shows the TM absorption subtracted by one half of the TE absorption, which approximately represents the absorption for the light polarization vertical to the wells. The spectra below 80 meV is not shown because of the low frequency cut off of the detector. (b) Calculated absorption coefficients of S3 at 15 K and 200 K based on the $k$-dependent 6-band **k·p** model with a half-width broadening of 15 meV. The top and the bottom panels show the absorption for the light polarized vertical and parallel to the wells, respectively.

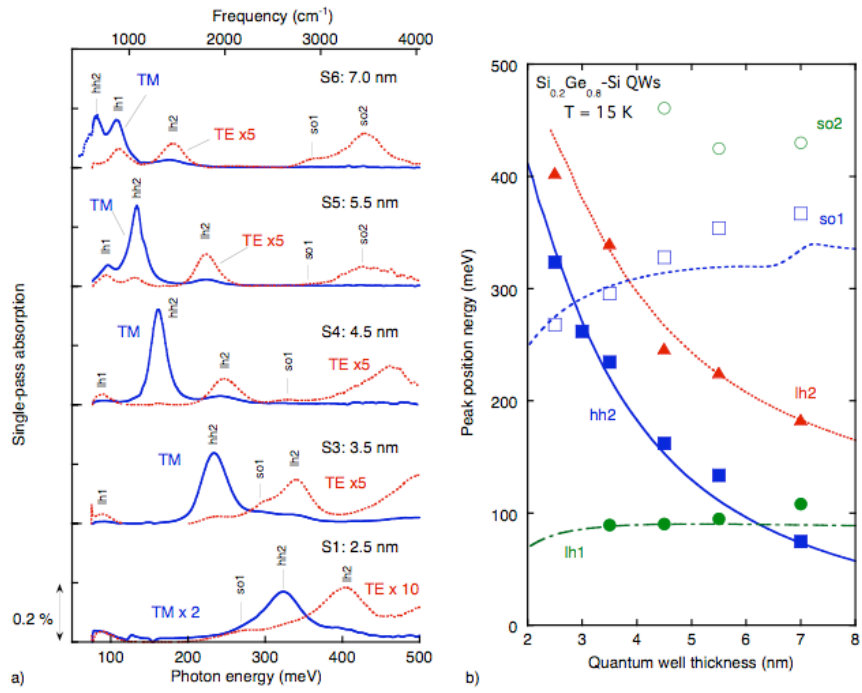

S. Tsujino et.al.: Figure 1

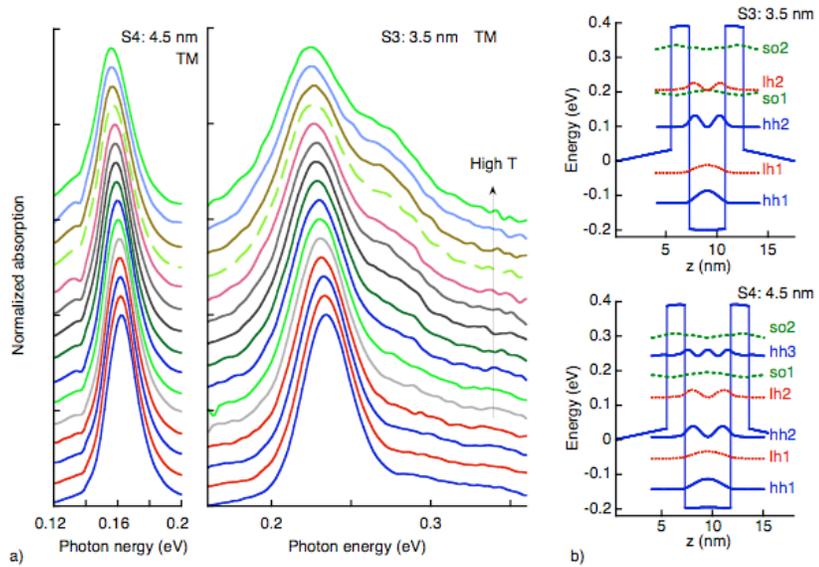

S. Tsujino et.al. : Figure 2

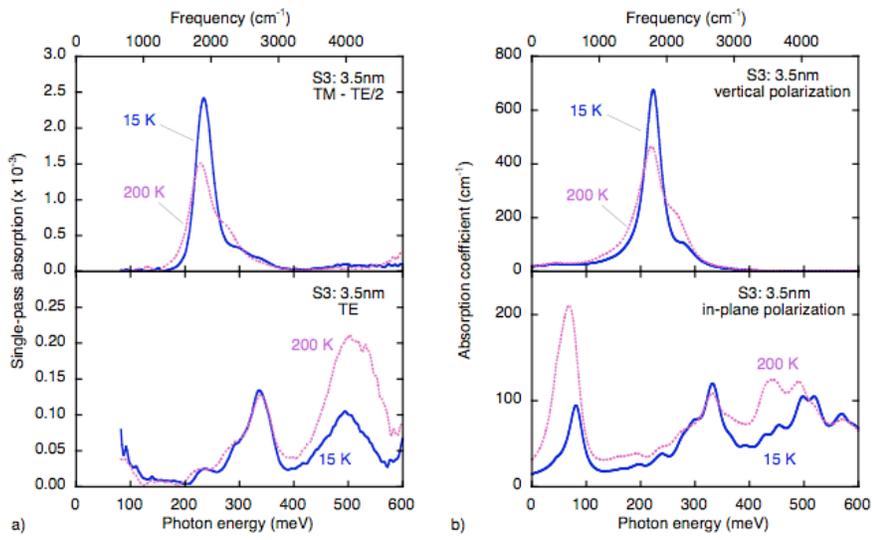

S. Tsujino et.al.: Figure 3

16